\documentclass[prc,letterpaper,notitlepage,nofootinbib,preprintnumbers,superscriptaddress]{revtex4-1} 

\usepackage[utf8]{inputenc}
\usepackage{amsmath,amssymb,amsfonts}

\usepackage{graphicx}

\usepackage{dcolumn}
\usepackage{bm}
\usepackage{color}
\usepackage{bbold}
\usepackage{url}

\usepackage{slashed}
\usepackage[svgnames]{xcolor}
\usepackage[english]{babel}
\usepackage{blindtext}
\usepackage{microtype}

\graphicspath{{images/}}

\usepackage[colorlinks=true,linkcolor=blue,urlcolor=blue,citecolor=blue]{hyperref}

\begin{document}
\title{Probing the gluonic structure of the nucleon through quarkonium production}

\author{Sylvester Joosten}
\affiliation{Temple University}

\date{\today}

\begin{abstract}
\noindent Heavy quarkonium production will provide for novel opportunities to study the gluonic structure of the nucleon in the near future.
Near threshold quarkonium production allows for direct experimental access of the dynamic origin of the nucleon mass as well as the nature of the color Van der Waals force,
while quarkonium production at high energies 
can be used to create a full three-dimensional tomographic image of the gluons inside the nucleon, constraining the gluonic radius of the nucleon.
\end{abstract}

\maketitle

\section{Towards an understanding of the nucleon mass}
The vast majority of the mass of the visible universe consists of atomic nuclei made up out of protons and neutrons.
At a more fundamental level, the nucleon mass can be seen as an emergent phenomenon from quantum chromodynamics (QCD), caused by the internal dynamics of the massless gluons, the almost massless quarks, and their color interaction.
In fact, modern calculations indicate that the proton mass is almost entirely unaffected by the value of the quark mass~\cite{Cloet:2013jya,Bhagwat:2003vw}.
This implies that the Higgs mechanism is largely irrelevant for `normal' matter.
The origin of the nucleon mass is a hot topic in nuclear science, highlighted in the 2015 long range plan~\cite{Geesaman:2015fha}: ``\textit{The vast majority [...] is due to quantum fluctuations of quark-antiquark pairs, the gluons and the energy associated with quarks moving around close to the speed of light.}''

Through \textit{ab initio} calculations, lattice QCD has made large strides to evaluate the mass of hadrons ~\cite{Durr:2008zz,Aoki:2008sm,Aoki:2009ix}. 
However, the origin of the nucleon mass itself in terms of its constituents remains elusive.
The difficulty lies in the fact that, unlike the proton spin, the proton mass does not lend itself to a frame independent decomposition.
In a covariant description~\cite{Kharzeev:1995ij}, the nucleon mass can be related to the trace of the energy-momentum tensor at zero momentum transfer. 
Here, the dominant contribution is proportional to the trace anomaly of QCD, implying that the nucleon mass can be considered to be the result of the vacuum polarization induced by the presence of the proton.
Alternatively, the nucleon mass can be written as the expectation value of the QCD Hamiltonian in the nucleon rest frame~\cite{Ji:1994av,Ji:1995sv,Lorce:2017xzd}. 
The proton mass then decomposes into contributions corresponding to the quark and gluon kinetic energy, the light quark masses, and a trace anomaly term $M_a$.

\begin{figure}[hbt]
\centering
\includegraphics[width=.38\textheight]{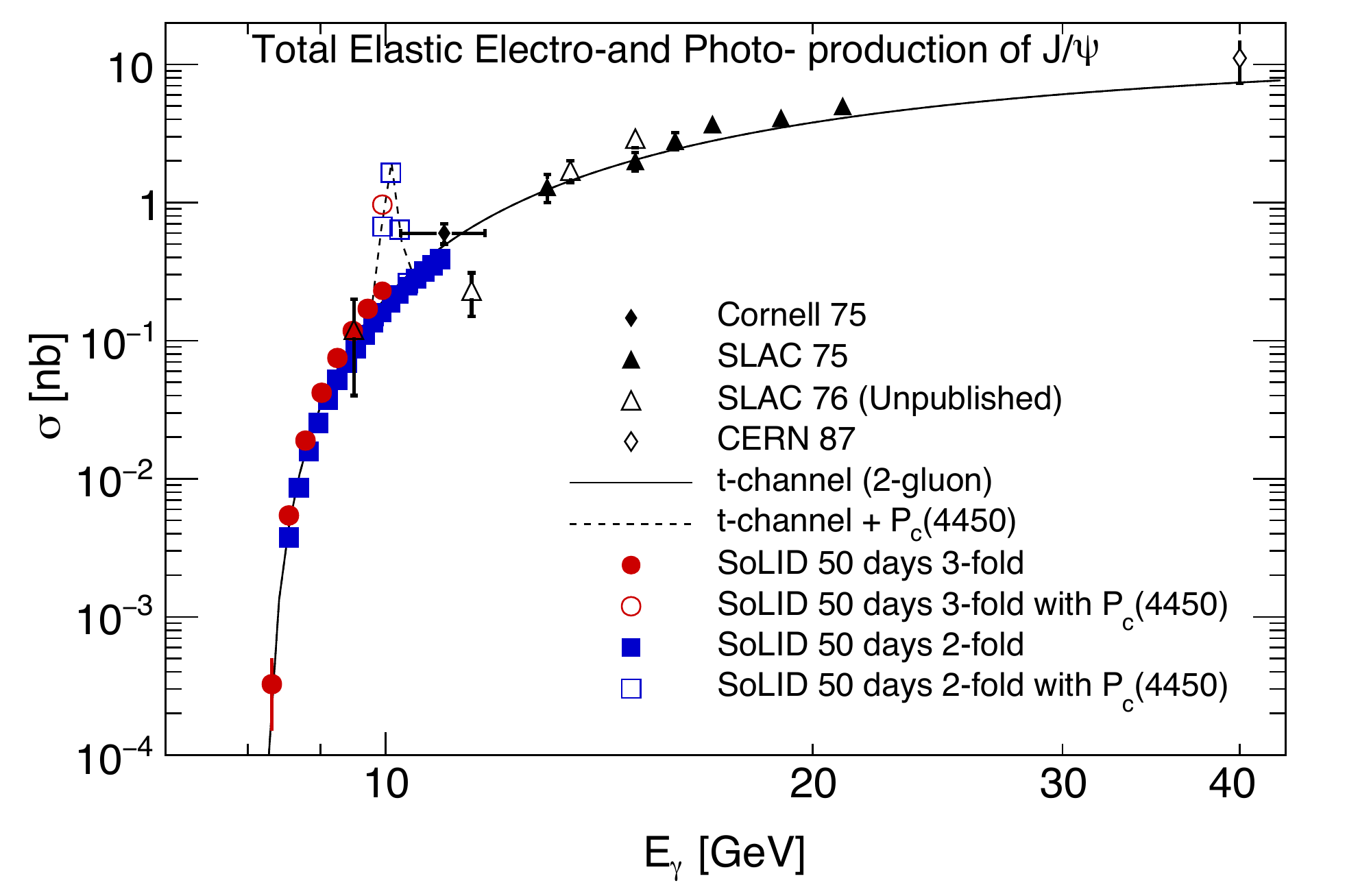}
\includegraphics[width=.265\textheight]{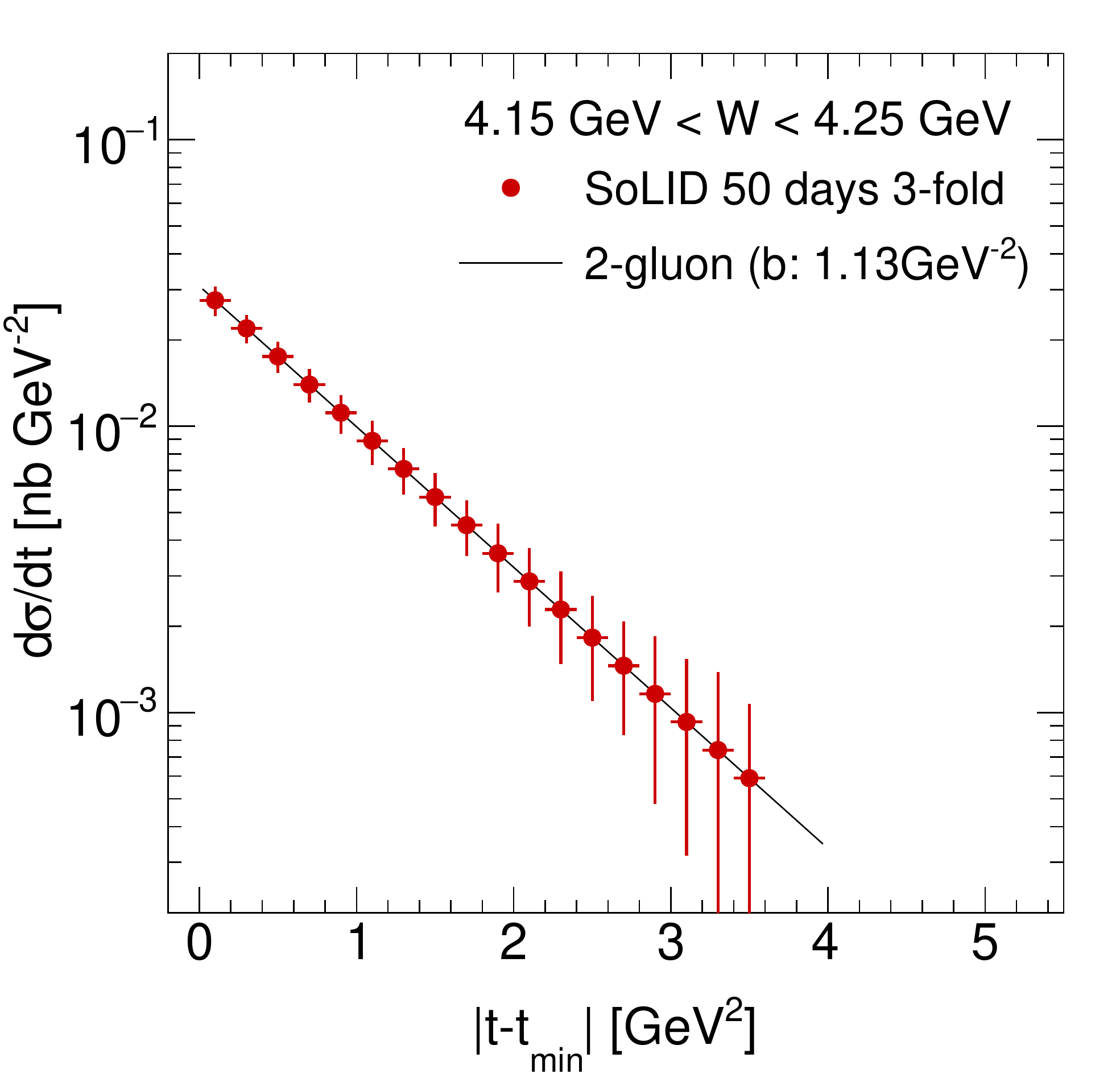}
\caption{
Left: Projected uncertainties of the total elastic $J/\psi$ electro (filled circles) and photo (filled squares)-production cross section using the SoLID detector, as a function of photon energy $E_\gamma$. 
Also plotted (open circles and squares) is the contribution of the larger mass LHCb pentaquark.
Right: differential cross section as a function of $ \vert t-t_{min}\vert$ in the case of electro-production in a $W$ bin very close to threshold, corresponding to the third electro-production point on the left figure. 
Figures taken from Ref.~\cite{Joosten:2018gyo}.
}
\label{solid-jpsi}
\end{figure}

It is clear that the trace anomaly is deeply related to the dynamic origin of the proton mass. 
It has not yet been determined experimentally, nor has it been constrained by a direct calculation in lattice QCD.
Experimental access to the trace anomaly is possible through quarkonium production on a proton close to the threshold~\cite{Kharzeev:1995ij,Kharzeev:1998bz,Gryniuk:2016mpk}.
In Jefferson Lab, experiment E12-12-006~\cite{SoLIDjpsi:proposal} will use the future solenoidal large acceptance device (SoLID) of Hall A to study this threshold region in detail. 
It will use a 3 $\mu$A beam on a 15 cm liquid hydrogen target for a total integrated luminosity of 43.2 ab$^{-1}$.
The projected results for the SoLID $J/\psi$ experiment are shown on Fig.~\ref{solid-jpsi}.
This experiment will provide for the high statistical power required to constrain the trace anomaly. 
Similarly, at an electron ion collider (EIC), the trace anomaly can be studied through elastic $\Upsilon$ production near threshold~\cite{Joosten:2018gyo}.
Combining the results from charmonium production at Jefferson Lab and bottomium production at an EIC will ensure redundancy while minimizing the theoretical systematic uncertainties.

\section{Color Van der Waals force and bound nucleon-quarkonium states}

The interaction between the color neutral quarkonium
and the nucleon is described by multiple gluon exchange. 
Unlike the case for nucleons in the nucleus, mesonic exchange channels are negligible because the nucleon and quarkonium do not share any valence quarks~\cite{Brodsky:1997gh}.
This purely gluonic force between color neutral objects is often referred to as the color Van der Waals force.

This force between quarkonium and the nucleon or nucleus is expected to be attractive, allowing for the existence of bound quarkonium-nucleon and quarkonium-nucleus states~\cite{Brodsky:1989jd}.
The sparsity of the available world data on $J/\psi$ production near threshold~\cite{Camerini:1975cy,Gittelman:1975ix,Anderson:1976sd,Barate:1986fq} has not prevented a long-running theoretical effort towards a quantitative grasp on this color Van der Waals force 
using techniques from non-perturbative and lattice QCD~\cite{Shevchenko:1996ch,Hayashigaki:1998ey,Yokokawa:2006td,Kawanai:2010ru,Tsushima:2011kh,Tsushima:2011fg}.
The most recent results from the NLQCD collaboration found a binding energy for $J/\psi$ in nuclear matter of 40 MeV or less~\cite{Beane:2014sda}, while a recent analysis in a phenomenological dispersive framework found a $J/\psi-p$ binding energy of $2.7\pm 0.3$ MeV~\cite{Gryniuk:2016mpk}.
The latter reference introduced a novel observable to constrain the quarkonium-nucleon scattering amplitude and binding energy through the measurement of the $\gamma p\rightarrow e^+e^-p$ forward-backward asymmetry in the vicinity of the $J/\psi$ resonant amplitude.
This asymmetry is caused by the interference between the Bethe-Heitler amplitude and the $J/\psi$ production amplitude.

The topic of the recently discovered $J/\psi-p$ resonances by the LHCb collaboration, commonly referred to as the LHCb pentaquarks~\cite{Aaij:2015tga}, is closely related to the existence of the color Van der Waals force.
The LHCb pentaquarks, which occur near the $J/\psi$ threshold, inspired many theoretical papers, e.g. 
Refs~\cite{Karliner:2015ina,Chen:2015loa,Guo:2015umn,Liu:2015fea,Wang:2015jsa,Kubarovsky:2015aaa,Karliner:2015voa,Eides:2015dtr,Chen:2016qju,Lu:2016nnt,Huang:2016tcr,Blin:2016dlf,Yamaguchi:2017zmn}.
However, without further measurements, the true nature of the observed states remains unclear.
On one hand, they can take the shape of either a five-quark state or a molecular bound 
state~\cite{Karliner:2015ina,Eides:2015dtr,Chen:2016qju,Lu:2016nnt,Huang:2016tcr}.
On the other hand, it is also possible the observations are due to kinematic enhancements~\cite{Liu:2015fea,Bai:2003sw}.
Real photo-production of $J/\psi$ near threshold is the ideal tool to distinguish between both explanations: a real $P_c$ state can be created through the $s$-channel and $u$-channel, while the type of kinematic enhancements that would play a role in the LHCb measurement do not occur in photo-production.

\begin{figure}[htb]
\centering
\includegraphics[width=.33\textheight]{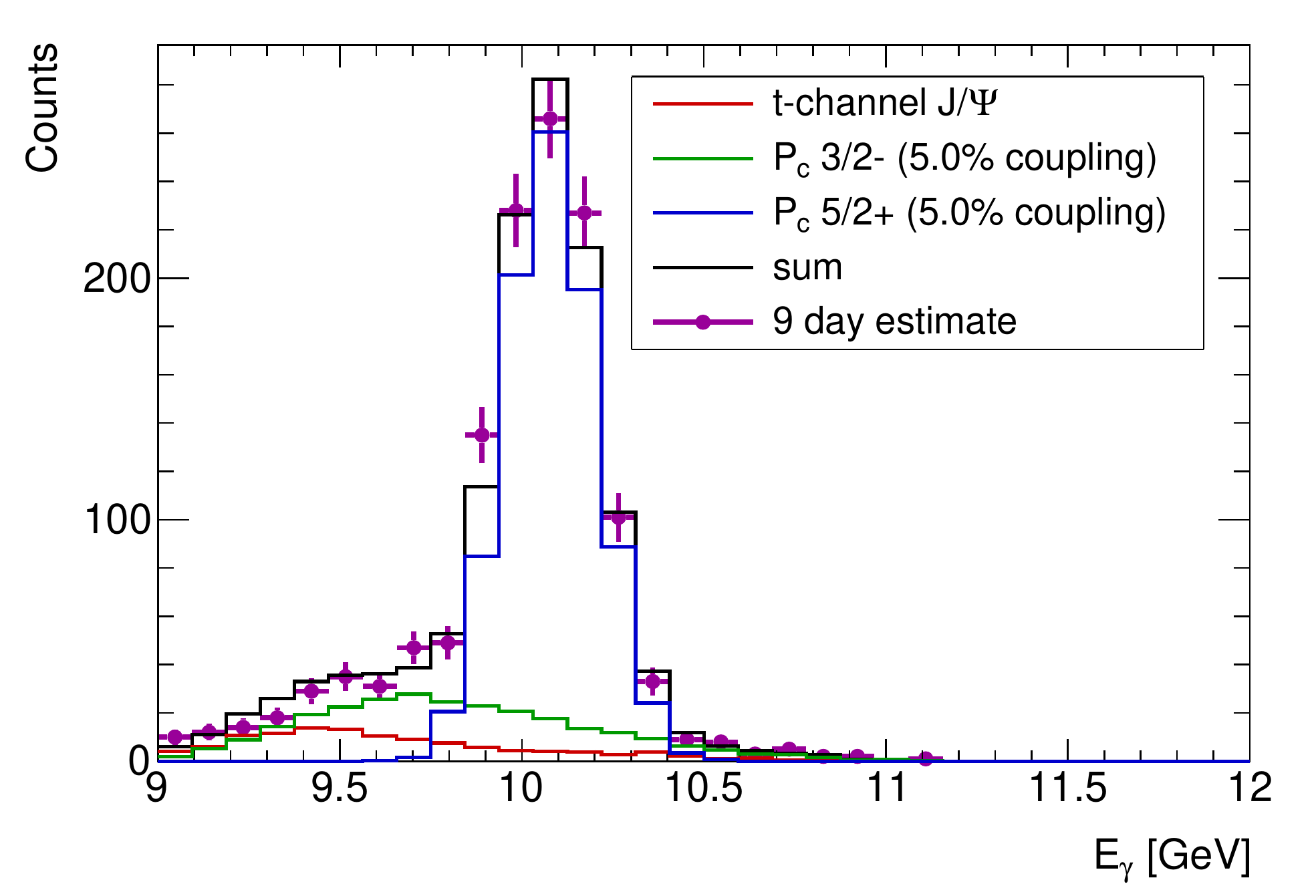}
\includegraphics[width=.33\textheight]{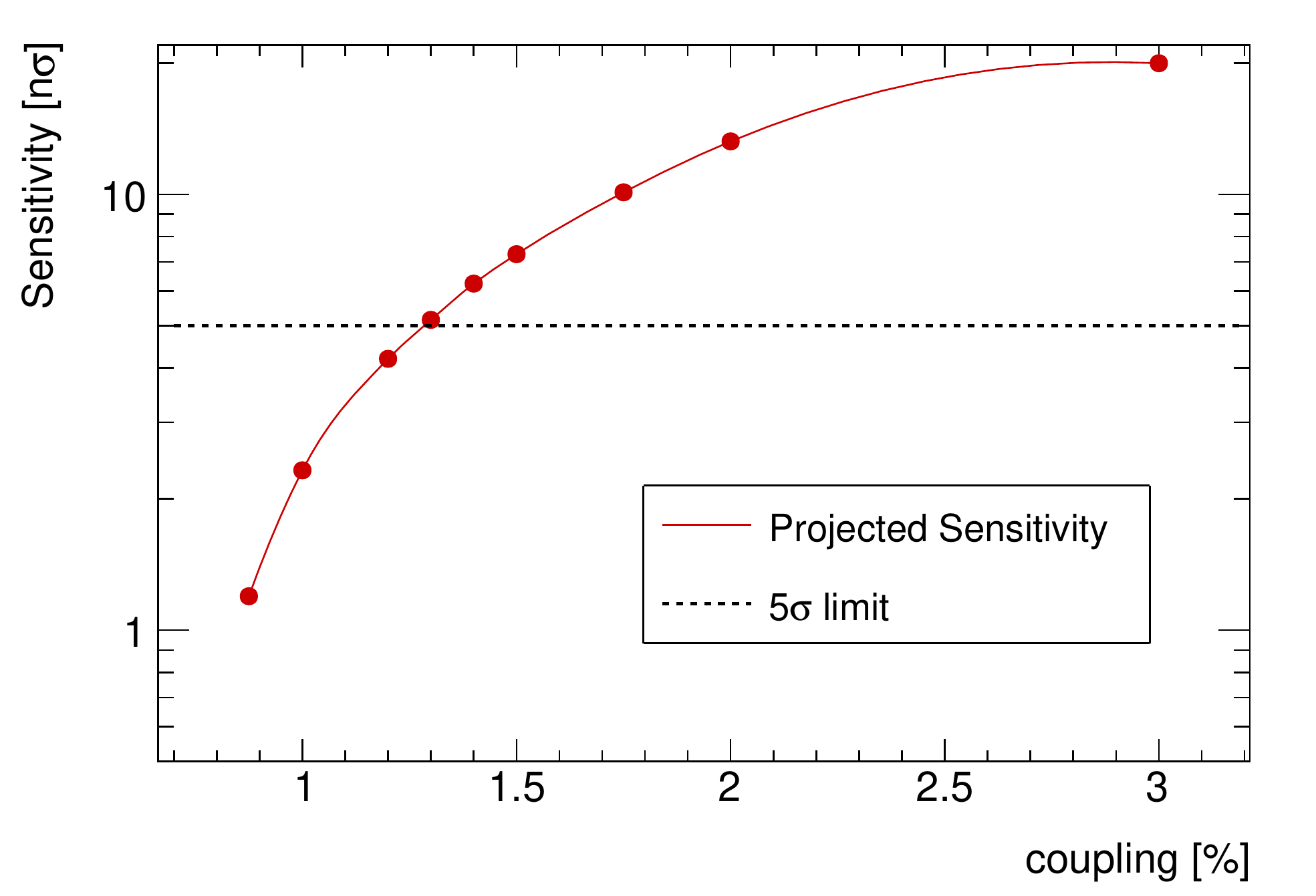}
\caption{
Left: The expected spectrum in a measurement of 9 days for the dedicated pentaquark-search experiment E12-16-007 in Hall C at Jefferson Lab,
assuming a $P_c$--to--$J/\psi$ $p$ coupling of 5\% according to the formalism of Ref.~\cite{Wang:2015jsa}.
Right: Sensitivity to the $P_c$ as a function of the coupling to the $J/\psi$ $p$ channel obtained from a log-likelihood analysis in this formalism.
The dashed line shows the $5\sigma$ level of sensitivity necessary for discovery. Figures taken from Ref.~\cite{Hafidi:2017bsg}.
}
\label{fig:pc}
\end{figure}

The new 12 GeV era at Jefferson Lab will provide crucial new measurements of the $J/\psi$ cross section near threshold.
The GlueX collaboration, as well as experiment E12-12-001A~\cite{CLAStcs:proposal,CLASjpsi:proposal} by the CLAS12 collaboration will provide the first new near-threshold data points in decades.
The dedicated pentaquark-search experiment E12-16-007~\cite{Meziani:2016lhg} in Hall C will use a high-intensity real photon beam to answer the question regarding the true nature of the LHCb pentaquark.
It will detect the decay $e^+e^-$ pair of the $J/\psi$ in both spectrometers of Hall C in two configurations: symmetric to measure the diffractive $t$-channel production (the `background') and asymmetric to measure the resonant production through the $s$-channel and $u$-channel (the `signal').
The projected results from this experiment are shown in Fig.~\ref{fig:pc}.
The SoLID $J/\psi$ experiment E12-12-006~\cite{SoLIDjpsi:proposal}, discussed in the previous section, will provide the necessary luminosity and acceptance to generate an ultimate multi-dimensional map of the threshold region.
Finally, one can speculate that, if the existence of bound $J/\psi$-p states is indeed confirmed, similar $\Upsilon-p$ states should also exist~\cite{Karliner:2015ina,Chen:2016qju,Yamaguchi:2017zmn,Karliner:2016joc}. This can be studied at an EIC~\cite{Joosten:2018gyo}.

\section{Gluon tomography and the gluonic radius of the proton}
The `shape' of the proton is currently understood in terms of its electromagnetic form factor and charge radius.
The precise gluonic structure is the missing component necessary to come to a full understanding of the actual matter distribution inside proton.
We do not yet know if we should imagine the gluons to be a spring-like structure at the heart of the proton, or rather a halo that reaches far beyond the edges of the traditional proton radius.

\begin{figure}[bth]
\centering
\includegraphics[width=.3\textheight]{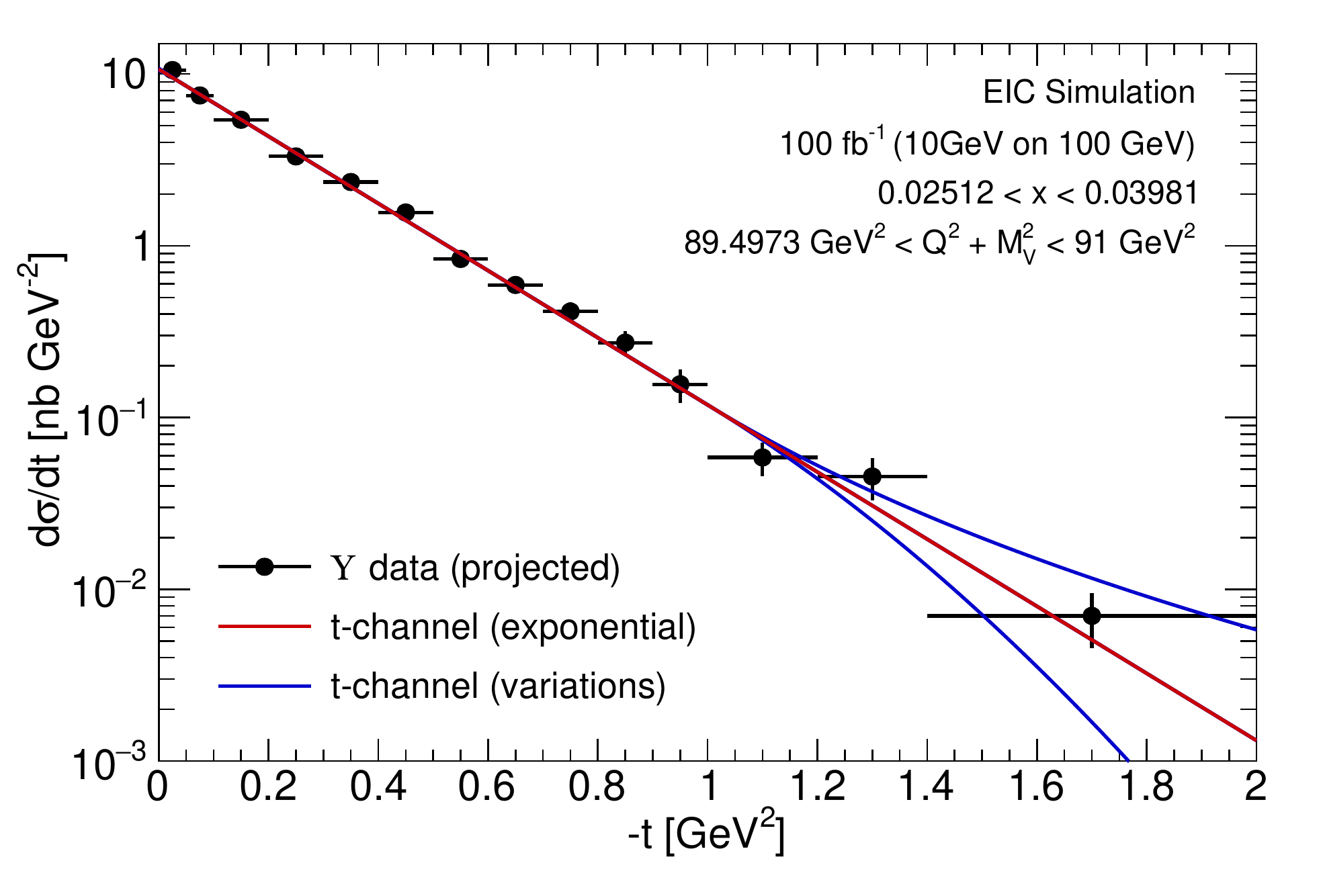}
\includegraphics[width=.4\textheight]{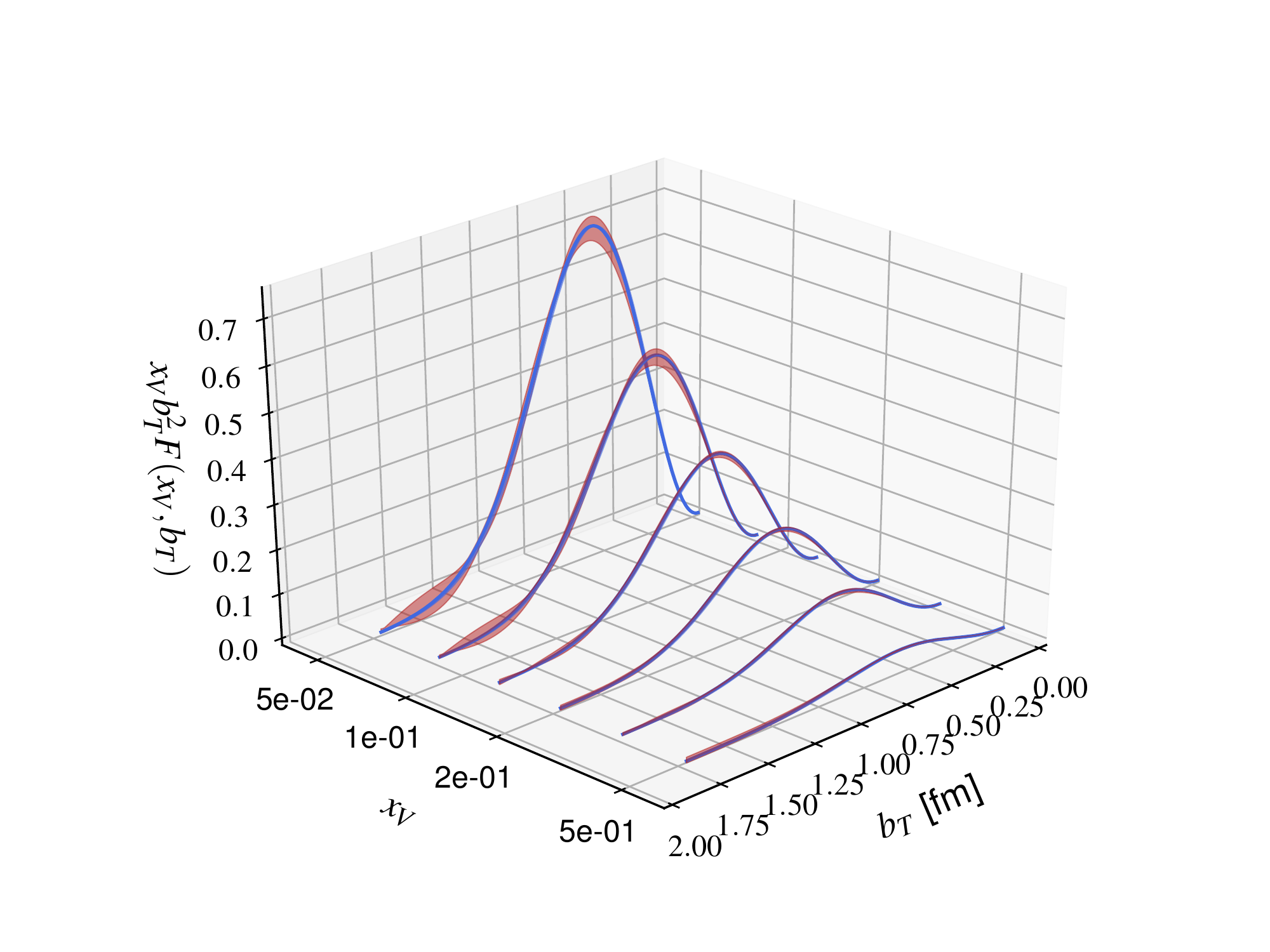}
\caption{Left: $t$-distributions from $\Upsilon$ elastic production for the lowest bin in $Q^2 + M_V^2$ and the lowest bin in $x_V$. The red line shows the expected exponential dependence of the cross section, and the blue lines show various different extrapolations for $t$ outside of the measured region. 
Right: Gluon GPD multiplied with $b_T^2$ as a function of the impact factor $b_T$ in six bins of $x_V$ for the lowest bin in $Q^2 + M_V^2$. This quantity highlights the precision at higher values of $b_T$, crucial for a precise determination of the gluonic radius. The blue band shows the statistical uncertainty of the fit, while the red band shows the statistical uncertainty added in quadrature with the systematic uncertainty due to the extrapolation at low and high $t$.
Figures taken from Ref.~\cite{Joosten:2018gyo}.}
\label{fig:upsilon-gpd}
\end{figure}

The white paper for the EIC~\cite{Accardi:2012qut} explains how the gluonic generalized parton distribution (GPD) that encodes the structure and dynamics of the gluons inside the nucleon can be accessed through $J/\psi$ electro-production at high energies.
The gluonic GPD can be directly related to the Fourier transform of the Mandelstam-$t$ distribution of the quarkonium production cross section~\cite{Burkardt:2002hr,Diehl:2002he,Boer:2011fh}.
In turn, this GPD can then be used to calculate the gluonic radius of the proton.
While it is not yet fully understood how next-to-leading order (NLO) corrections play into this extraction, they are expected to be non-negligible.
The study of $\Upsilon$ electro-production can be used to control for this uncertainty.
Due to the much larger mass of the $b$-quark, the higher order corrections will be significantly suppressed compared to those for the $J/\psi$.
Fig.~\ref{fig:upsilon-gpd} show a projection for the extracted gluon GPD through exclusive $\Upsilon$ electro-production assuming a 10 GeV electron bean colliding with a 100 GeV proton beam for a total integrated luminosity of 100 fb$^{-1}$.
The figures shows a single bin in $Q^2 + M_\Upsilon^2$, which is the relevant resolution scale for meson production. It is binned in slices of $x_V=(Q^2 + M_V^2)/(2P\cdot q)$, which replaces the standard Bjorken variable $x_\text{B}=Q^2/(2P\cdot q)$.
The precision of this measurement will be more than sufficient to complement the profiles obtained through the $J/\psi$ measurements and will provide for an important test of the universality of the extracted gluonic GPD when NLO corrections are accounted for.

\section{Conclusion}

Jefferson Lab in the 12 GeV era will measure the $J/\psi$ electro- and photo-production cross section near threshold with
an unprecedented precision.
It will investigate some of the most fundamental open questions in QCD: What is the origin of the proton mass? And what role does the trace anomaly play? What is the strength of the color Van der Waals force in the quarkonium-nucleon system? What is the nature of the LHCb charmed pentaquark?
Three-dimensional nucleon tomographic imaging will be possible at the proposed EIC through $J/\psi$ electro-production at high energies. 
It will also provide access to new quarkonium observables through the $\Upsilon$ production, both near threshold and at high energies.
The heavy $\Upsilon$ will be a complimentary probe to the threshold $J/\psi$ program at Jefferson Lab as well as the $J/\psi$ tomography program at the EIC, trading statistical precision for smaller theoretical uncertainties.
These measurements, together with advances in lattice QCD calculations and phenomenology, will greatly advance our understanding of QCD, and in particular the curious role of gluons inside the proton.

\begin{acknowledgments}
I thank the organizers for the opportunity to present this work, supported in part by the U.S. Department of Energy Grant Award
DE-FG02-94ER4084.
\end{acknowledgments}

\bibliography{paper}

\end{document}